# The two body problem: proprioception and motor control across the metamorphic divide

Sweta Agrawal[1*], John C Tuthill[1]

[1]Department of Physiology and Biophysics, University of Washington, Seattle, Washington, USA
*Correspondence and lead contact: sagrawal@uw.edu

**Abstract**

Like a rocket being propelled into space, evolution has engineered flies to launch into adulthood via multiple stages. Flies develop and deploy two distinct bodies, linked by the transformative process of metamorphosis. The fly larva is a soft hydraulic tube that can crawl to find food and avoid predators. The adult fly has a stiff exoskeleton with articulated limbs capable of long-distance navigation and rich social interactions. Because the larval and adult forms are so distinct in structure, they require distinct strategies for sensing and moving the body. The metamorphic divide thus presents an opportunity for comparative analysis of neural circuits. Here, we review recent progress toward understanding the neural mechanisms of proprioception and motor control in larval and adult *Drosophila*. We highlight commonalities that point toward general principles of sensorimotor control and differences that may reflect unique constraints imposed by biomechanics. Finally, we discuss emerging opportunities for comparative analysis of neural circuit architecture in the fly and other animal species.

**Highlights**

- We compare strategies used by *Drosophila* larvae and adults to sense and move the body
- Sensorimotor circuits of larval and adult flies are adapted for their distinct modes of locomotion
- Developmental lineages are an effective means to identify neuronal cell-types within and across animals
- Comparing larval and adult fly nervous systems may provide general insight into brain evolution

**Introduction**

The adolescence of a fruit fly is relentless. A fly egg is laid soon after fertilization (Fig. 1) [1,2]. Within 24 hours, neural progenitor cells, called neuroblasts, will have produced ~15,000 neurons that will wire together into circuits that sense and move the body of an alert, mobile tube of cells: the 1st instar larva [2]. For the five serene days of its youth, the larva will explore its environment and do its best to pack on the milligrams. All the while, neuroblasts continue to divide, generating clusters of immature, postmitotic neurons that will remain dormant until adulthood [3,4]. Less than 120 hours after birth, the cuticle of the larva will harden into a protective shell, a puparium, as the larva prepares to transition to adulthood. For the next four days, the fly will gut and remodel its entire body plan, metamorphosing from a squishy tube to a stout hexapod with a recognizable head, legs, and wings. The nervous system that controlled the larval tube is disassembled; some parts are discarded, some re-used, and some are built from scratch to form a much larger nervous system (~150,000 neurons) [3,5]. About 10 days into its life, the fly will emerge from the puparium as a full-figured adult. It is ready to take its first nip of yeasty apple juice, launch itself into the air, and fly away to fulfill its biological imperative.

The nervous systems of the adult and larval fly are in some respects similar, while in other ways unique. Many of the familiar anatomical structures in the adult fly brain – the optic lobes [6], olfactory system [2], and mushroom bodies [7] – are also recognizable in the larvae. In some cases, they have been found to exhibit similar architecture, though the larval circuits contain far fewer total cells [7]. By comparison, it is more challenging to recognize superficial structural similarities among the diffuse sensorimotor circuits of the fly ventral nerve cord (VNC), the invertebrate analog of the spinal cord [8].

This may be partly because the body forms of larvae and adults are so distinct. Their bodies are not only shaped differently, but they also exhibit distinct motor behaviors. To move forward, larvae use a system of crawling locomotion that opposes the movement of muscles against a hydraulic skeleton, while adults use jointed limbs for walking and flight. Therefore, we might expect some of the largest differences between adult and larval flies to occur in the circuitry for proprioceptive sensing and motor control.

Perhaps more than any other sense, proprioception is intimately tied to the body: proprioceptive sensing is influenced by the biomechanical properties of the structures in which they are embedded [9] and the firing patterns of proprioceptors must be formatted to control the body they inhabit. For example, limbed animals sense strains concentrated at joints and transmit this information to motor neurons that control limb muscles. On the other hand, soft-bodied animals sense tensile forces within the body surface and transmit this information to motor neurons that control contraction of the body. Despite these differences, many of the basic building blocks for proprioceptive sensing are fundamentally similar.

In this review, we begin by comparing and contrasting the proprioceptive sensors used by fly larvae and adults to monitor their bodies. We then consider how these proprioceptive signals are transformed by the nervous system and used for feedback control of movement, such as during locomotion. Finally, we discuss exciting new opportunities for comparative analysis within sensorimotor circuits of the fly and speculate about general insights that may be gleaned from bridging the metamorphic divide.

**Major fly proprioceptors**

Proprioception relies on mechanosensory neurons (proprioceptors) embedded in joints, muscles, and cuticle throughout the body [10]. These sensory neurons create an internal representation of body state by monitoring body kinematics including joint angles, joint stresses and strains, muscle length, and muscle tension. Mechanosensory neurons that directly detect mechanical forces generated by the external world are referred to as exteroceptors. However, the line between exteroceptors and proprioceptors is often blurry. Exteroceptors can be stimulated during self-generated movement, and proprioceptors can be stimulated when external stimuli cause body parts to move [11].

The majority of larval proprioceptors are multidendritic, non-ciliated (Type II) neurons that tile the body wall (Fig. 1B), whereas the majority of adult proprioceptors are ciliated (Type I) neurons, such as campaniform sensilla, chordotonal organ (ChO) neurons, and hair plates, that attach to cuticular structures that sense the movement and position of joints (Fig. 1C) [12]. Larvae also possess some Type I ChO neurons that may function as both proprioceptors and exteroceptors, detecting muscle stretch as well as sound, vibration, and gentle touch [13]. Knocking out or mutating mechanotransduction channels in ChO neurons of fly larvae produces abnormal crawling and changes in the frequency of bending, suggesting that ChO neurons are important for regulating locomotion [14–17]. However, recent studies looking at the development of sensorimotor circuits in the *Drosophila* embryo demonstrated that disrupting mechanosensory activity of ChO neurons during spontaneous muscle contractions of the developing embryo affects their connectivity with their post-synaptic partners and disrupts the maturation of motor circuits [18,19]. Thus, it is unclear if such locomotor defects are a result of these neurons' proprioceptive function during crawling or their critical role in circuit development.

Proprioceptors can vary in their mechanical sensitivity and stimulus tuning, even within a single limb segment or muscle. For example, a recent study by Mamiya et al., showed that the femoral chordotonal organ (FeCO) of the adult fly leg is composed of genetically distinct proprioceptor subclasses that detect and encode distinct kinematic features of the femur-tibia joint, including tibia position, directional movement of the tibia, or tibial vibration [20]. The cell bodies of each of these proprioceptor subclasses, denoted *hook*, *club*, and *claw*, reside in separate parts of the FeCO in the leg, and their axons project to distinct regions of the fly VNC. In the larva, a pair of recent studies used high-speed fluorescence imaging to visualize changes in the structure and activity of multidendritic proprioceptors in the larval body wall during crawling [21,22]. The resulting videos reveal the dynamic flexing of these neurons' dendrites during different phases or directions of segment contraction: when larvae crawl forward, the most posterior dendrites of the ventral class I neuron (vpda) and the dorsal class I neuron, ddaE, fold inward during each contraction, followed by bending of the dendrites of the more anterior dorsal class I neuron (ddaD). When larvae move backward, the dendrite of the ddaD neurons are the first to fold inward. Proprioceptor activity correlates with their dendritic deformations. Interestingly, though they are morphologically similar, the bipolar neurons on the ventral and dorsal side of each abdominal segment (dbd and vbd) are activated by stretch and contraction, respectively [22]. This result suggests that the dynamics of dendritic deformations cannot completely explain proprioceptor tuning and that some other mechanism must underlie dbd and vpd proprioceptor responses during crawling.

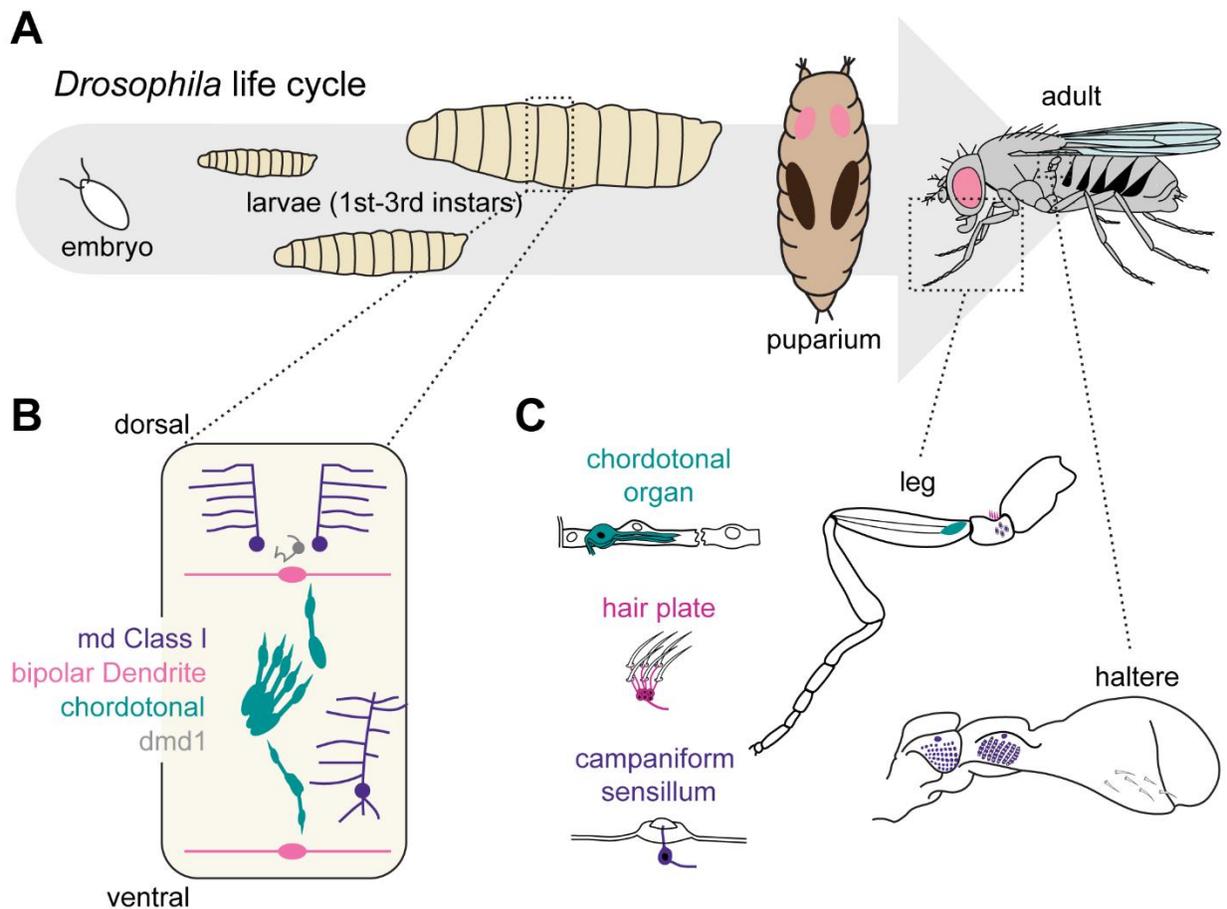

Figure 1. The proprioceptive system of *Drosophila*. A) Schematic of the *Drosophila* life cycle, progressing from the egg (left), through three larval instars, a puparium, and finally the adult fly (right). B) The major proprioceptive sensory neurons in the larva include the multidendritic (md) Class I neurons, dorsal and ventral bipolar dendrite neurons, the chordotonal (ChO) neurons, and dmd1. Modified from Vaadia et al., (2019). C) Left, schematics of three major proprioceptor types found in the adult, including the chordotonal organ, hair plate, and campaniform sensillum. Right, schematics of the leg and haltere, with the locations of a subset of proprioceptors. Note that these examples do not represent the full complement of leg or haltere proprioceptors, but rather illustrate examples of each type of sensor.

The unique sensitivity of each proprioceptor subtype is the result of cell-intrinsic biophysical properties and the biomechanics of the tissues in which the neuron is embedded. *Drosophila* adults and larvae mechanosensory neurons express many of the same mechanotransduction channels, including TRP channels like NOMPC, Piezo channels like Piezo1 and Piezo2, degenerin/epithelial sodium (DEG/ENaC) channels, and DmTMC, which is a member of the transmembrane channel-like (TMC) protein family [12,23–25]. How these channels affect the dynamics of proprioceptors' response to mechanical perturbation is not well established. A major challenge for probing the function of these channels is the limited accessibility of most fly mechanosensory neurons, which are small and often tightly wrapped in connective tissue and supporting cells. As a result, recording transduction currents from primary mechanosensory neurons in the fly is often extremely difficult. Instead, most evidence comes from indirect methods, such as measurements of mechanically evoked calcium signals or patch-clamp recordings from downstream interneurons [12]. Methods like single-cell RNA-seq [26] will also be helpful for understanding cell-intrinsic properties that determine proprioceptor stimulus sensitivity.

Structural adaptations, such as the biomechanical properties of the neuron's attachment to a joint, can also determine the tuning of mechanosensory neurons. For example, the decrease in stiffness along the mammalian cochlea promotes frequency selectivity along its length [27]. In the adult fly, white noise analysis suggests that campaniform sensilla (CS) are all fairly similar: wing CS, regardless of their location, can be described by predictive models using only two stimulus features, one of which approximates the derivative of the other [28]. Instead, structural adaptations may determine their stimulus sensitivity, such as the shape of their cuticular cap, viscoelastic properties of the neuron's coupling to the cap, and their location relative to sources of cuticular strain [29–31]. Biomechanical modelling [32] may help clarify the relative roles of neuron-intrinsic and structural adaptations that determine proprioceptor stimulus sensitivity.

**Proprioceptive control of movement**

Motor behaviors in the fly require precise and coordinated neural control of dozens to hundreds of muscles. This coordination is mediated by populations of motor neurons, which translate commands from the central nervous system into dynamic patterns of muscle contraction. Proprioceptors provide feedback to motor circuits to modulate posture, stabilize the body, coordinate the timing of phase transitions, and refine motor output [10].

Neuromuscular systems can be large and complex: several muscles often control the movement about a single joint, and each of those muscles may be innervated by several motor neurons. How does the nervous system coordinate the activity of these motor neuron populations to flexibly control the force, speed, and precision of body movements? A recent study by Azevedo et al., found that motor control of the adult *Drosophila* leg is streamlined according to a hierarchical control model known as the size principle [33]. The size principle, which was originally formulated to explain motor neuron recruitment in cats, posits that motor neurons are recruited in an order that depends on the amount of force they produce [34]. Motor neurons controlling slow or weak movements are typically recruited first, followed by motor neurons that control progressively stronger and faster movements [34]. A key assumption of the size principle is that all motor neurons within a motor pool receive similar presynaptic input, and the likelihood that a motor neuron fires in response to that input is due to neuron-intrinsic properties like its axon diameter and input resistance. Azevedo et al., found that leg motor neurons of the adult fly, like that of the cat [34], exhibit a coordinated gradient of anatomical, physiological, and functional properties that establish a hierarchical recruitment order (Fig. 2A-C). However, while control of the femur-tibia joint is largely consistent with the size principle, the dynamics of proprioceptive feedback varied across the different motor neurons (Fig. 2D). Thus, control of the adult *Drosophila* leg motor system is more complex than a straightforward implementation of the size principle.

The larva has a segmented body, with each segment containing ~30 bilateral pair of motor neurons as well as 30 bilateral body wall muscles that form 'spatial muscle groups' based on common location and orientation (Fig. 2E) [35,36]. A larva predominantly locomotes by crawling: both forward and backward crawling are symmetric movements achieved by propagation of muscular contraction and relaxation along the body in a segmentally coordinated manner. In an effort to understand whether the same motor neurons and muscles within a segment are active during both forward and backward locomotion, Zarin et al., utilized pan-muscle activity imaging [36]. They found that all motor neurons and their target muscles in each segment are active during both forward and backward locomotion, but six of those muscles demonstrate a notable change in their recruitment timing: the three muscles in the VO spatial muscle group and muscles 2, 11, and 18 (each in a different spatial muscle group). When they used electron microscopy (EM) reconstruction to map the major inputs to larval motor neurons, they found that while some premotor interneurons innervated specific spatial muscle groups, most of them innervated multiple motor neurons across spatial groups. In addition, no premotor interneurons selectively innervated the co-active motor neurons that were differentially recruited during backward versus forward crawling. Thus, many of the same muscles, motor neurons, and premotor neurons within a segment are likely involved in both forward and backward crawling. Instead, a recent study by Kohsaka et al., suggests that two pairs of intersegmental feedback neurons that target a common group of premotor neurons coordinate the appropriate contraction of muscles by acting as delay circuits representing the phase lag between segments (Fig. 2F) [37]. Each pair of intersegmental neurons, Ifb-Fwd and Ifb-Bwd, are active during only forward or backward crawling, respectively. Thus, these neurons coordinate activation of synergistic muscles across body segments in a distinct manner for forward and backward crawling.

Basic crawling behavior in larvae does not require proprioceptive feedback; animals can still crawl even when synaptic release is blocked in mechanosensory neurons [14]. When sensory feedback is surgically removed, the larval CNS still endogenously produces segmentally coordinated motor output [38]. However, eliminating proprioceptive input leads to qualitative changes in the crawling gait [38,39]: peristaltic waves are slower, and body segments contract twice as much. Larvae often display reduced crawling speeds, increased head cast frequencies, and enhanced backward locomotion. Adult flies that lack functional leg proprioceptors show deficits in posture, walking speed, stance duration, as well as more variable foot placement [11,40]. Additionally, adult flies lacking proprioceptive feedback have a harder time compensating for injury like the loss of a leg [41]. In both larvae and adults, evidence suggests that proprioceptive feedback is important for coordinating motor transitions inter- and intra-segmentally, or between joints and across multiple legs [35,42,43]. Challenging the proprioceptive system by incorporating variable and uneven terrain into laboratory experiments may reveal additional functions of sensory feedback during locomotion.

In addition to walking, adult flies have a second mode of locomotion: flight. Walking and flight operate via distinct sets of limbs controlled by distinct muscles that operate on different timescales (30 steps/s compared to 200 wingbeats/s) [44]. Additionally, the organization of the VNC and patterns of descending neuron innervation from the brain suggest that neural control of these two locomotor behaviors are mostly independent [45]. While wings and legs possess the same types of proprioceptive sensory neurons, their relative distributions differ – flight-associated appendages possess hundreds of CS (~80 on each wing and ~140 on each haltere) and fewer than 100 ChO neurons, whereas each leg has about 150 ChO neurons and under 50 CS neurons [46,47]. That CS are so much more numerous on flight appendages than legs may reflect unique strategies to detect the different forces that must be sensed during the two behaviors, for example ground contact forces during walking versus aerodynamic forces during flight.

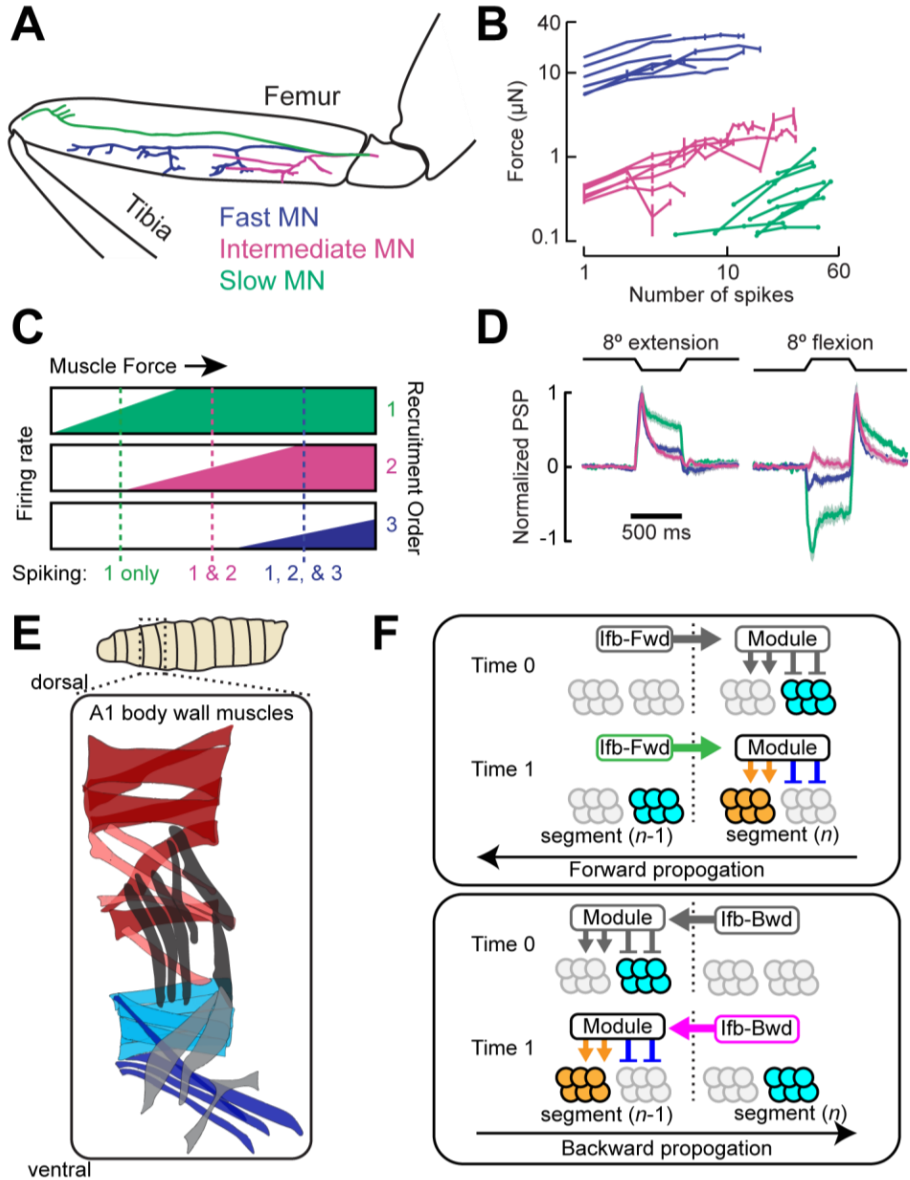

Figure 2. Motor control strategies in the adult and larva. A) Schematic showing the axons of fast (blue), intermediate (magenta) and slow (green) motor neurons and where they innervate muscles in the femur of the adult leg. These motor neurons control flexion of the femur-tibia joint. B) Peak average force vs. number of spikes for fast (blue), intermediate (magenta), and slow (green) motor neurons. The number of spikes in slow neurons is computed as the average number of spikes during positive current injection steps minus the baseline firing rate; that is, the number of additional spikes above baseline. C) Schematic illustrating firing rate predictions of the recruitment hierarchy. As force increases, left to right, first the slow neuron begins firing (green line), then the intermediate neuron (magenta line). Once the fast neuron spikes, both intermediate and slow neurons are already firing (blue dotted line). D) Time-course of normalized average responses (± s.e.m. of baseline subtracted responses) to tibia extension (left) and flexion (right, n = 7 fast, seven intermediate, nine slow cells). A-D modified from Azevedo et al., (2020). E) Schematic of the 29 muscles of larval abdominal segments (A1) from external view. Spatial muscle groups are indicated by color. Modified from Zarin et al., (2019). F) Model of the roles of intersegmental feedback neurons Ifb-Fwd (top) and Ifb-Bwd (bottom) in transforming the progression of a wave front to the intersegmental delay via a conserved circuit module in forward (top) and backward (bottom) propagations. The circles represent either the transverse (orange) or longitudinal (blue) muscle groups. Modified from Kohsaka et al., (2019).

The density of proprioceptors is particularly high on the haltere, a mechanosensory organ unique to flies. Halteres are modified hind wings that no longer provide aerodynamic lift – instead, they function as small, club-shaped mechanosensory organs that detect inertial forces during flight via an array of CS mechanosensory neurons distributed at the base [48]. Haltere CS fire action potentials on each cycle of the haltere's oscillation, which beat up and down in anti-phase to the wingstroke. These action potentials are precise and reliable over a broad range of haltere oscillation frequencies [49]. Lateral displacements of the haltere, such as might happen during instabilities due to wind gusts or wing damage, can recruit specific CS afferents or phase shift their activity, enabling the nervous system to monitor self-motion in a manner analogous to the vertebrate vestibular system [50]. Feedback from wing and haltere CS influences the firing of wing and neck motor neurons on a wingstroke-to-wingstroke basis [51,52]. Interestingly, though it no longer functions as a wing, the haltere still possesses a set of steering muscles at its base [53]. A recent study by Dickerson et al., demonstrated that these muscles are active during flight and receive descending visual input regarding global visual rotations about the body's cardinal axes [53]. They posit that visual information tunes the strength of haltere mechanosensory feedback during flight by slightly adjusting the haltere's motion via activation of these muscles. Thus, haltere sensory neuron activity may reflect a combination of visual and mechanosensory information. Behavioral experiments show that flies are able to execute corrective flight maneuvers with remarkably low latency – flies subjected to an applied torque from a magnetic field acting on a dorsally tethered metal pin recover from this perturbation within five wingbeats [54]. Control-theoretic flight simulations suggest that, to maintain stability, flies must sense their body orientation every wingbeat [55].

Beyond being used directly for stabilizing motor control, proprioceptive information is important for motor planning and navigation. The central complex, a set of neuropils in the central insect brain, processes multisensory information from the environment and integrates it with information about the insect's internal state to guide motor output [56]. Neural activity within the adult central complex is predictive of walking speed and turning behavior even in the absence of visual input, suggesting that it receives proprioceptive input from the legs [57]. Additionally, the activity of some central complex neurons correlates with movement of the antennae or the haltere [58]. Finally, the central complex has been implicated in leg motor control, and electrical stimulation of central complex neurons can alter walking behavior [59,60]. However, almost nothing is known about how proprioceptive information is conveyed to the central complex, mostly because little is known about the information encoded more generally by ascending neurons projecting from the VNC to the brain. There are several more ascending than descending axons in the neck connective of the adult fly [61]. However, compared to descending neurons, little is currently known about the anatomy and function of ascending neurons, including neurons that convey mechanosensory information to brain circuits in the adult [62].

*Drosophila* larvae do not have a recognizable central complex until the third instar. Even then, it lacks terminal branches and synapses, suggesting that it is not yet functional [63]. Third instar larvae from genetic strains with abnormal adult central complexes demonstrate locomotor defects, though it is unclear if this is due to the absence of a functional central complex [64].

**Central transformation of proprioceptive information**

Central neurons within the VNC transform proprioceptive sensory information so that it can be used by the brain or motor circuits to adjust behavior. Though the larval and adult life stages look and behave differently, their VNCs are generated by the same segmental array of ~30 neuroblasts [65]. Each neuroblast divides to form an A (Notch$^{ON}$) and B (Notch$^{OFF}$) hemilineage, characterized by the absence or presence of Notch signaling (Fig. 3A) [66]. Developmental lineages are an effective means to classify neuronal cell-types: neurons within a hemilineage are morphologically similar, innervate similar axon tracts, express the same transcription factors, and release the same primary neurotransmitter in larvae and adults (with the exception of neurons born during early embryonic neurogenesis) [4,65,67,68]. However, it remains an open question if neurons within a hemilineage serve similar functions across the metamorphic divide. There are currently only a few instances of individual neurons that are known to persist from larval to adult stages and serve similar functions, including individual motor and sensory neurons and mushroom body neurons [3,69]. For example, the core MDN-Pair1 interneuron circuit, which controls backward walking, persists through metamorphosis and serves a similar function in adults and larvae, despite the fact that its upstream and downstream connectivity is profoundly remodeled [70]. Beyond a few specific examples, though, we currently lack effective methods to comprehensively track individual neurons through metamorphosis. Due to changes in gene expression, genetic driver lines do not always label the same cells in larvae and adults [71]. Because hemilineages can be reliably identified in the larval and adult nervous systems, this classification scheme can provide a convenient, though relatively coarse, means of identifying cell types and comparing their connectivity and function across metamorphosis.

In adults, neurons from several hemilineages – 8A, 8B, 9A, 10A, 13B, and 19A – are downstream of leg proprioceptive sensory neurons [72,73]. Recording their activity with *in vivo* whole-cell electrophysiology revealed that some of these central neurons integrate information across proprioceptor subtypes to construct multimodal sensorimotor representations (Fig. 3B). For example, 9Aa neurons receive input from all three types of FeCO sensory neuron types (hook, club, and claw), resulting in these neurons encoding a complex combination of tibial flexion, tibia position, and high frequency tibia vibration. In contrast, 13Ba neurons receive input from just a single type of FeCO sensory neuron and encode extended tibia positions. Both neurons are involved in reflexive control of joint position: optogenetic activation of 9Aa neurons in headless flies causes extension of the femur-tibia joint, whereas optogenetic activation of 13Ba neurons causes flexion of the femur-tibia joint. The significance of the specific representations of femur-tibia joint kinematics encoded by these central neurons remains an open question, one that may be answered by using EM reconstruction to trace proprioceptive signals through intermediate circuits to the motor neurons that control leg joint movement.

In the larval VNC, Even-Skipped+ neurons (lineage 8) are part of a sensorimotor circuit that maintains left-right symmetry of muscle contraction amplitude in *Drosophila* larvae [74]. Basin neurons (lineage 9) relay multimodal mechanosensory and proprioceptive input to motor circuits [75,76]. A study by Valdes-Aleman et al., used "comparative connectomics," combining EM reconstruction, functional imaging of neural activity, and behavioral experiments to understand how central neurons, such as the Basin neurons, integrate multimodal signals from their presynaptic mechanosensory partners [19]. When Valdes-Aleman et al., genetically shifted the location of axonal projections of mechanosensory neurons, they discovered that the majority of postsynaptic partners all redirected their dendritic projections towards the shifted mechanosensory input, thereby retaining synaptic connections; only one new partner was gained and only weakly connected partners were lost. However, the number of synapses made by mechanosensory neurons onto central neurons changed enough that it impaired mechanosensory behavior. This result suggests that most connections between mechanosensory neurons and their postsynaptic partners are determined by genetically specified, partner-derived cues. When the synaptic output of mechanosensory neurons was silenced throughout development, Valdes-Aleman et al. observed an increase in the number of synapses onto excitatory postsynaptic partners and a decrease in those onto inhibitory postsynaptic partners, indicating that neuronal activity plays a key role in circuit development. This unique study illustrates the potential of using comparative connectomics to understand the development and function of proprioceptive circuits.

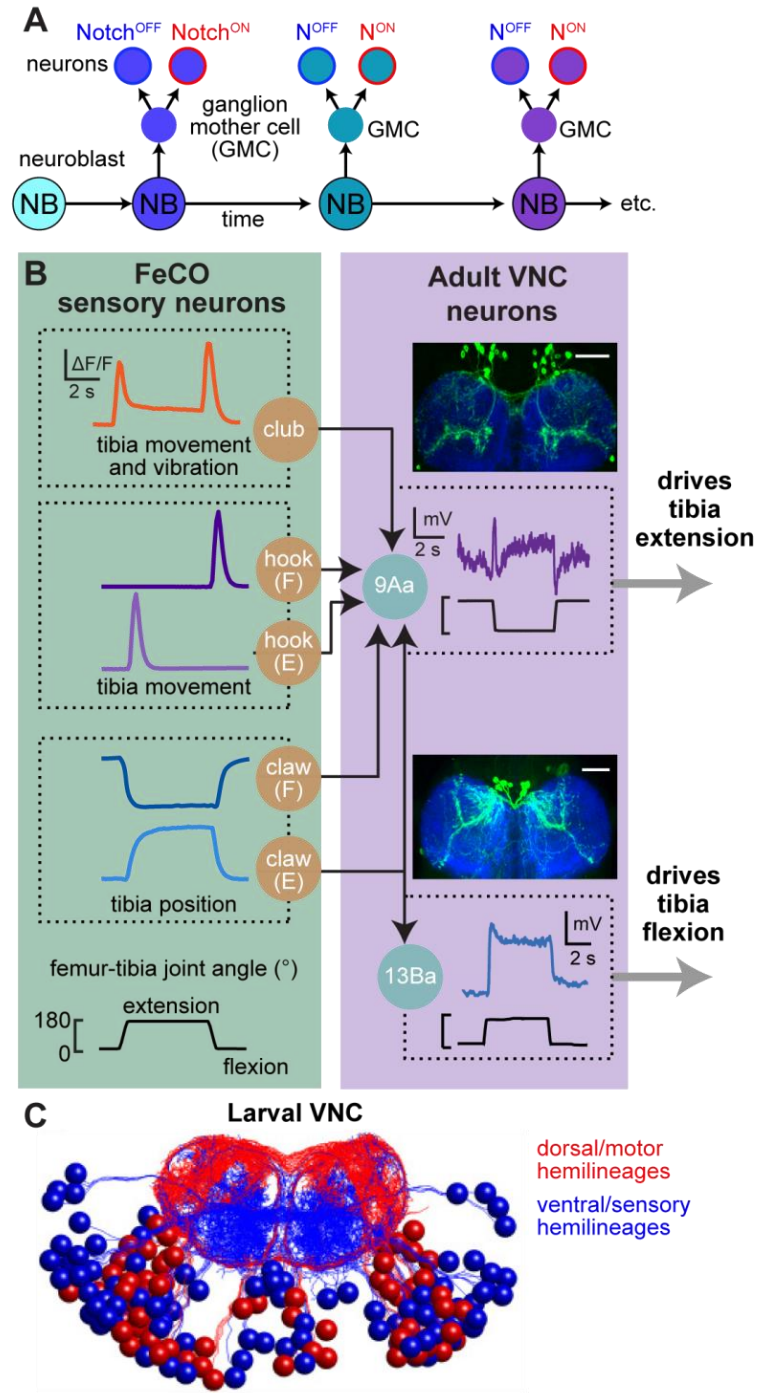

Figure 3. A) Neuroblasts (NB) undergo asymmetric cell division, producing another NB and a ganglion mother cell (GMC). Hemilineages are formed by GMC asymmetric division into a pair of post-mitotic neurons; during this division, the Notch inhibitor Numb is partitioned into one Notch$^{OFF}$ (N$^{OFF}$, blue) neuron and one Notch$^{ON}$ (N$^{ON}$, red ) neuron. B) Sensory encoding of femur-tibia joint movement and position by femoral chordotonal organ (FeCO) sensory neurons and the downstream central neurons 9Aa (formerly known as 9Aα) and 13Ba (formerly known as 13Bα). Sensory neuron activity was recorded via *in vivo* GCaMP6f fluorescence (ΔF/F), whereas central neuron activity was recorded via whole-cell patch clamp electrophysiology (mV). Inset images are confocal images of the anatomy of each cell type in the prothoracic VNC, scale bar: 25 µm (green: GFP, blue: nc82). Optogenetic activation of 9Aa neurons causes femur-tibia joint extension, whereas optogenetic extension of 13Ba neurons causes femur-tibia joint flexion. Data taken from Mamiya et al., (2018) and Agrawal et al., (2020). C) Superimposition of all dorsal candidate hemilineages (N$^{ON}$, red) and all ventral candidate hemilineages (N$^{OFF}$, blue). Modified from Mark et al., (2021).

A recent study by Mark et al., sought to understand connectivity trends more broadly across several larval VNC hemilineages [77]. By mapping the developmental origin and connectivity of 80 bilateral pairs of larval central neurons from seven different neuroblasts, they found that hemilineages innervate sensory or motor neuropils in a Notch-dependent manner: The Notch$^{ON}$, or A hemilineage, always projected to the dorsal/motor neuropil, whereas the Notch$^{OFF}$, or B hemilineage, always projected to the ventral/sensory neuropil (Figure 3C). They also demonstrated that hemilineage neurons that were born in a similar temporal cohort share similar connectivity and synapse location. Currently, it is unknown if adult VNC neurons follow a similar pattern. However, with the recent acquisition of a complete electron microscopy volume of the VNC [61] and innovations in tracing the parent lineage of neurons in the adult VNC [71], it may soon be feasible to address these questions. Such comparisons of hemilineage function and connectivity between larvae and adults will yield insight into how central circuits for proprioception are repurposed following metamorphosis.

**Looking forward**

One of the most remarkable and counterintuitive aspects of the fly's life cycle is that the larval body plan evolved more recently than the adult [78]. The earliest insects, relatives of modern bristletails and silverfish, did not have a larval form. Instead, the juvenile body plan more closely resembled the adult. Modern hemimetabolous insects, like cicadas, grasshoppers, mantises, and dragonflies, also adhere to this developmental trajectory. To accommodate changes in size, they progress through a series of molts as they grow. As a result, their means of proprioceptive sensing, central processing, and motor control do not require extensive remodeling across developmental stages, with the exception of the emergence of flight in adults. Holometabolous insects, like *Drosophila*, evolved a lifestyle that requires a massive overhaul of the body and nervous system, perhaps to allow a single individual to take advantage of multiple niches over its lifetime. The recent work on the function and development of proprioceptive and motor control circuits in *Drosophila* provides a unique opportunity to understand the evolution of holometaboly by connecting the animal's two distinct life stages. It is also a unique opportunity to understand the degree to which the function or structure of an animal's proprioceptive and motor systems are adapted to their particular body plan.

Our current opinion is that the most fascinating extant question in neuroscience is how neural circuits have evolved to produce the incredible diversity of animal behavior found on Earth. Here, we suggest that larval and adult *Drosophila* can serve as a testbed for developing comparative molecular and structural approaches to investigate the diversity and evolution of neural circuits. As EM connectomics, single-cell transcriptomics, and gene-editing become cheaper and easier, these techniques can be applied to investigate the nervous systems of more diverse insect species. In this endeavor, conserved hemilineage identity may provide a Rosetta Stone for identifying and comparing neuronal cell types. In the long run, as tools for identifying developmentally-related cell-types improve, this approach may also be extended to other branches of the animal kingdom.

**Acknowledgements**
We are grateful for insightful comments on the manuscript from Jim Truman, Brandon Mark, Ellen Lesser, and Ellie Heckscher. SA is supported by a National Institute of Health grant K99NS117657. JCT is a New York Stem Cell Foundation – Robertson Investigator and was additionally supported by the Searle Scholar Program, the Pew Biomedical Scholar Program, the McKnight Foundation, and National Institute of Health grants R01NS102333 and U19NS104655.

**Recommended reading**
Papers of particular interest, published within the period of review, have been highlighted as:

\* of special interest
\*\* of outstanding interest

**\*\*Carreira-Rosario et al. [18]:** This study uses calcium imaging to record spontaneous neural activity in developing *Drosophila* embryos. They find that this spontaneous activity is highly stereotyped across embryos and important for the normal maturation of motor circuits. Inhibition of muscle contractions or mechanosensory input from chordotonal organ neurons results in premature and excessive spontaneous activity and aberrant larval locomotion.

**\*\*Valdes-Aleman et al. [19]:** This unique study illustrates the potential of using comparative connectomics to understand the development and function of proprioceptive circuits. They genetically shifted the projection pattern of mechanosensory axons or silenced the synaptic output of mechanosensory neurons, and then used EM reconstruction, functional imaging of neural activity, and behavior experiments to understand how such manipulations altered neural connectivity.

**\*\*He et al. [21]:** Similar to Vaadia et al. [22], this paper used high-speed fluorescence imaging to visualize changes in the structure and activity of multidendritic proprioceptors in the larval body wall during crawling. Proprioceptive neurons in animals mutant for TMC showed a near-complete loss of movement related calcium signals, suggesting that TMC channels may be activated by membrane curvature of the dendrites of body wall proprioceptors.

**\*\*Vaadia et al. [22]:** Similar to He et al. [21], this paper used high-speed fluorescence imaging to visualize changes in the structure and activity of multidendritic proprioceptors in the larval body wall during crawling. During cycles of segment contraction and extension, different proprioceptor types exhibited sequential activity related to the dynamics of each neuron's terminal processes. As a result, larval body wall proprioceptors encode a continuum of position during all phases of crawling.

**\*Sun et al. [24]:** Using electron tomography, this study found that fly campaniform sensilla contain thousands of force-sensitive ion channels that are aligned to the intracellular microtubule cytoskeleton. Mechanical modeling suggested that the pattern is structurally and functionally optimized such that force-sensitive channels are located at regions that are subject to large activating forces, thereby enhancing the sensitivity and broadening the dynamic range of mechanosensation.

**\*\*Azevedo et al. [33]:** This study found that *Drosophila* leg motor neurons exhibit a coordinated gradient of anatomical, physiological, and functional properties similar to hierarchical motor neuron recruitment in vertebrates (the size principle). However, the dynamics of proprioceptive feedback varied across the different motor neurons, suggesting that control of the joint is more complex than an implementation of the size principle.

**\*Zarin et al. [36]:** This study characterized larval muscle activity patterns and premotor/motor circuits to understand how they generate forward and backward locomotion. They found that all body wall motor neurons (MNs) are activated during both behaviors, but a subset change recruitment timing for each behavior. They then used EM to reconstruct a full segment of all 60 MNs and 236 premotor neurons (PMNs) to identify PMN-MN circuit motifs that could all contribute to generating distinct behaviors.

**\*\*Kohsaka et al. [37]:** This study identified two pairs of higher-order premotor excitatory interneurons that intersegmentally provide feedback to the adjacent neuromere during motor propagation. The two feedback neuron pairs are differentially active during either forward or backward locomotion but commonly target a group of premotor interneurons that provide excitatory inputs to transverse muscles and inhibitory inputs to the antagonistic longitudinal muscles.

**\*Dickerson et al. [53]:** This study found that visual input during flight modulates haltere muscle activity, thereby altering the mechanosensory feedback from the haltere. This result suggests that rather than acting solely as a gyroscope to detect body rotation, halteres also function as an adjustable clock to set the spike timing of wing motor neurons.

**\*Phelps et al. [61]:** This study used serial-section electron microscopy (EM) to acquire a synapse-resolution dataset of the ventral nerve cord of an adult female *Drosophila*. Using this dataset, they then reconstructed all 507 motor neurons that control the limbs. They also provide open access to the dataset and reconstructions registered to a standard atlas to permit matching of cells between EM and light microscopy data.

**\*\*Lacin H [67]:** This study identified the embryonic neuroblast origin of the adult neuronal lineages in the ventral nerve cord via lineage-specific GAL4 lines and molecular markers. Their lineage mapping revealed that neurons born late in the embryonic phase show axonal morphology and transcription factor profiles that are similar to the neurons born post-embryonically from the same neuroblast.

**\*Lee and Doe [70]:** This study found that Pair1, an interneuron that receives input from the moonwalker descending neuron (MDN) and inhibits forward crawling in the larva, persists through metamorphosis into the adult, where it continues to inhibit forward locomotion. Thus, the MDN-Pair1 neurons are an interneuronal circuit that generates similar locomotor behavior at both larval and adult stages.

**\*\*Chen et al. [72]:** By combining optogenetic activation and 2-photon calcium imaging, this study maps the functional connectivity between leg proprioceptors and downstream neurons in the adult fly. They then verify these connections via EM reconstruction.

**\*\*Agrawal et al. [73]:** This study investigates how second-order neurons in the Drosophila ventral nerve cord process proprioceptive information from the fly leg. Optogenetic stimulation paired with kinematic analyses reveals that different subtypes of second-order neurons mediate diverse behaviors, including reflexive control of limb posture.

**\*Mark B et al. [77]:** This study mapped the developmental origin of 160 interneurons from seven bilateral neuroblasts and identified them in an EM reconstruction of the Drosophila larval central nervous system. They found that lineages concurrently build the sensory and motor neuropils by generating sensory and motor hemilineages in a Notch-dependent manner.